\documentclass[aps,prl,superscriptaddress,amsfonts,amsmath,amssymb,showpacs,floatfix,reprint,longbibliography,footinbib]{revtex4-2}

\usepackage{times}
\usepackage{url}
\usepackage{bm}
\usepackage{graphicx}
\usepackage{amsmath}
\usepackage{amstext}
\usepackage{amssymb}
\usepackage{amsfonts}
\usepackage{amsbsy}
\usepackage{verbatim}
\usepackage{color}
\usepackage[colorlinks=true, urlcolor=blue, linkcolor=blue, citecolor=blue, pdftex]{hyperref}
\usepackage[capitalise]{cleveref}
\usepackage{multirow}
\usepackage{floatrow}
\usepackage{float}
\usepackage{tikz}
\usepackage{gensymb}
\usepackage{textcomp}
\usepackage{enumitem}
\usepackage[version=3]{mhchem}
\usepackage{afterpage}
\usepackage{pbox}
\usepackage{makecell}
\usepackage{dsfont}
\usepackage{siunitx}
\usepackage{fancyhdr}
\usepackage{stackengine,wasysym,scalerel}
\usepackage{latexsym}
\usepackage{cancel}
\usepackage{array, multirow, bigdelim, makecell, booktabs}
\usepackage[misc]{ifsym}
\usepackage{physics}

\begin{document}

\title{Polaronic correlations from optimized ancilla wave functions for the Fermi-Hubbard model}
  
\author{Tobias M\"uller}
\affiliation{Institut f\"ur Theoretische Physik und Astrophysik and W\"urzburg-Dresden Cluster of Excellence ct.qmat, Julius-Maximilians-Universit\"at W\"urzburg, Am Hubland, Campus S\"ud, W\"urzburg 97074, Germany}
\author{Ronny Thomale}
\affiliation{Institut f\"ur Theoretische Physik und Astrophysik and W\"urzburg-Dresden Cluster of Excellence ct.qmat, Julius-Maximilians-Universit\"at W\"urzburg, Am Hubland, Campus S\"ud, W\"urzburg 97074, Germany}
\affiliation{Department of Physics and Quantum Center for Diamond and Emergent Materials (QuCenDiEM), Indian Institute of Technology Madras, Chennai 600036, India}
\author{Subir Sachdev}
\affiliation{Department of Physics, Harvard University, Cambridge MA 02138, USA}
\author{Yasir Iqbal}
\affiliation{Department of Physics and Quantum Center for Diamond and Emergent Materials (QuCenDiEM), Indian Institute of Technology Madras, Chennai 600036, India}

 \date{\today}

\begin{abstract}
We employ a family of ancilla qubit variational wave-functions [Zhang and Sachdev, \href{https://doi.org/10.1103/PhysRevResearch.2.023172}{Phys. Rev. Res. 2, 023172 (2020)}] to describe the polaronic correlations in the pseudo-gap metal phase of a hole-doped 2D Fermi-Hubbard model. Comparison to ultra-cold atom quantum simulator data [Koepsel {\it et al}., \href{https://doi.org/10.1126/science.abe7165}{Science 374, 82 (2021)}] reveals both qualitative and quantitative agreement with the numerical analysis from half-filling up to 80\% hole-doping, capturing the crossover from the polaronic regime to the Fermi liquid observed around $40\%$ doping.
\end{abstract}

\maketitle

{\it Introduction.---}A variational wave-function {\it Ansatz} attempts to grasp the core essence of a given phase or parameter regime through the correlation and entanglement profile assumed by the principal wave-function structure, while reminiscent wave-function parameters are then fixed through the variational principle imposed on energy expectation values built from a given microscopic Hamiltonian. Facing a task as intricate as the nature of electronic order and criticality in the pseudogap regime of high-$T_c$ cuprates immediately poses challenges to reaching a level of description where such a variational {\it Ansatz} could even be formulated: (i) Being {\it sui generis} at finite doping, the wave-function {\it Ansatz} has to cover spin and charge excitations, and to account for their inter-dependencies; (ii) the pseudogap metal and the conventional Fermi liquid metal parent states are crucially different states with distinct quasiparticles, yet the {\it Ansatz} needs to provide a comprehensive connectivity between these limits.  

In the spin liquid perspective on the pseudogap \cite{LeeRMP}, parton theories yield mean-field Hamiltonians of charge neutral spin $S=1/2$ spinons and charge $e$ spin $S=0$ holons, and corresponding wave functions can be formulated for such states. It has long been recognized \cite{SS94,WenLee96,Stanescu_2006,Berthod_2006,Rice_2012,KaulKim07,Sakai_2009,Qi_2010,Mei_2012,Punk_2015,Chiu_19,Grusdt_19,Skolimowski_2022} that the charge doping, within such a fractionalized quasiparticle picture, is reconciled through spinon-holon bound states which have the same quantum numbers as electrons. A systematic theory of a finite density of such bound states in a phase without antiferromagnetic long-range order was recently provided by the ancilla qubit method \cite{zhang2020}. It provides an {\it Ansatz} which promises an adequate starting point for a reconciliation of the pseudogap regime from a variational principle. Formed by a physical Hubbard model layer coupled to an ancilla double-layer, it offers a sufficient amount of complexity to create all needed limits to navigate the pseudogap parameter regime. In the pseudogap metal (FL*) limit, an entangled physical and first ancilla layer enables the formation of small effective hole pockets, while the second ancilla layer equips the {\it Ansatz} with the fractional spin excitations needed by the violation of the Luttinger volume \cite{MO00,TSMVSS04,APAV04}. In the Fermi liquid metal (FL) limit, the physical layer can readily form large hole pockets, and does not require additional employment of the ancilla layers. Either way, the ultimate projection to the physical subspace can eventually be enforced by triggering ancilla local rung-singlet formation from a sufficiently large fictitious Heisenberg coupling within the ancilla double layer. Instead of performing such a projection immediately, one accounts for the expanded Hilbert space through a gauge theoretical assignment of the ambiguity implied by the artificially added ancilla layers. This is a recurrent motif in several classes of variational wave-functions, such as likewise implemented through projected mean-field states for spin liquids, where the projection is given by the Gutzwiller exclusion of the fermionic double occupancy while the mean-field states are characterized and categorized by the gauge group structure~\cite{Wen-2002}.   

As much as it would be in principle a worthwhile undertaking to now confront doping-dependent experimental evidence on cuprates with such a universal variational approach, it would almost certainly be destined to fail. This is because disorder, finite temperature, doping-dependent screening functions, and, in particular, sample sensitivity unavoidable for chemical doping would provide an overwhelming variability that might render it an impossible task to compare variational results at different dopings. This is crucially alleviated within a cold-atom quantum simulator. Hubbard-type Coulomb repulsion can be fixed independent of doping, disorder is negligible, and microscopic snapshots allow for a precise experimental measurement of individual correlation components. 

In this paper, we perform an extensive variational Monte Carlo simulation of the repulsive Fermi-Hubbard model on a square lattice with an ancilla qubit {\it Ansatz}, and compare against the experimental evidence provided by the cold-atom quantum simulator data read out of Koepsell {\it et al.\/}~\cite{koepsell2021}. 

{\it Model and variational wave function.---}
To connect to this data, we study the paradigmatic repulsive Hubbard model on the square lattice
\begin{equation}
    H = -t\sum\limits_{\langle i,j\rangle,\sigma = \uparrow,\downarrow} c^\dagger_{i,\sigma} c_{j,\sigma} + U \sum\limits_{i} c^\dagger_{i,\uparrow} c_{,\uparrow} c^\dagger_{i,\downarrow} c_{i,\downarrow} \label{eq:hubbardham}
\end{equation}
setting $U/t = 7.4$ and use a finite size cluster of $10\times10$ sites, matching the interactions and dimensions of the experimental cold-atom setup~\cite{koepsell2021}. To model the ground state wave funtion of \cref{eq:hubbardham}, we use the ancilla qubit {\it Ansatz} presented in Ref.~\cite{zhang2020}. Its essence is to augment the physical electrons by two hidden layers of localized qubits, i.e., spins with $S = \frac{1}{2}$, the first of which is directly coupled to the electronic degrees of freedom, while the second only interacts locally with the first via an antiferromagnetic Heisenberg interaction. The Heisenberg interaction is taken to be large, thereby effectively removing the magnetic background introduced in this construction by forcing the spins in the ancilla layers to form local singlets. By choosing a spin-liquid state on the second layer, this ancilla construction allows for the introduction of non-trivial effective spin-correlations in the physical electron system.

The variational wave-function is then the ground state of a mean-field Hamiltonian describing a layer of itinerant fermionic holes $d$ and two layers of ancilla spins $\mathbf{S}_{1/2}$. The latter are treated in terms of Abrikosov pseudo-fermions, leading to a representation given by $\mathbf{S}_{1/2} = \sum\limits_{\sigma, \sigma'}  {f_{1/2}}^\dagger_{i,\sigma} \mathbf{\tau}_{\sigma, \sigma'} {f_{1/2}}_{i,\sigma'}$~\cite{Abrikosov-1965}. To make this mapping exact, an additional constraint
\begin{equation}
    \sum\limits_\sigma {f_{1/2}}^\dagger_{i,\sigma}{f_{1/2}}_{i,\sigma} = 1 \quad \forall~i\,, \label{eq:spinconst}
\end{equation}
i.e., enforcing half-filling of the pseudo-fermions per site is enforced.

The advantages of this fractionalization scheme are twofold: firstly, it allows for a direct coupling between $d$ and $f_1$ by means of a hybridization term. Secondly, one can rely on the well-understood parton construction of spin-liquids to choose a suitable Hamiltonian for $f_2$. For the itinerant and first ancilla layer, we employ
\begin{equation}
\begin{split}
 \mathcal{H}_{c,f_1} = &- t\sum\limits_{\langle i,j\rangle,\sigma}  d^\dagger_{i,\sigma} d^{}_{j,\sigma} +  t_{f_1} \sum\limits_{\langle i,j\rangle ,\sigma}  {f_1}^\dagger_{i,\sigma} {f_1}^{}_{j,\sigma} \\&+ \Phi \sum\limits_{i,\sigma} (d^\dagger_{i,\sigma}{f_1}^{}_{i,\sigma} + {f_1}^\dagger_{i,\sigma}{d}^{}_{i,\sigma}) \\&- \mu_d \sum\limits_{i,\sigma} d^\dagger_{i,\sigma} d^{}_{i,\sigma} -\mu_{f_1} \sum\limits_{i,\sigma} {f_1}^\dagger_{i,\sigma} {f_1}^{}_{i,\sigma} \label{eq:mfham}
 \end{split}
\end{equation}
as the mean-field Hamiltonian. To fix the overall energy scale, we choose the itinerant hopping as the reference, setting $t=1$. The chemical potentials $\mu_d$ and $\mu_{f_1}$ are chosen, such that the $f_1$ layer is half-filled and simultaneously the density in the $d$ layer is $1+p$, such that $p$ is the hole-doping fraction of the physical system. Additionally, numerical investigation of the quantum geometric tensor reveals, that the hopping $t_{f_1}$ and hybridization $\Phi$ are not independent parameters, leaving us with a single variational degree of freedom, which we choose to be $\Phi$ by setting $t_{f_1} = 1$.

For the second ancilla layer, we choose a mean-field Hamiltonian having a pairing term with $d$-wave symmetry
\begin{equation}
    \begin{split}
     \mathcal{H}_{f_2} = &-  t_{f_2} \sum\limits_{\langle i,j\rangle ,\sigma}  {f_2}^\dagger_{i,\sigma} {f_2}^{}_{j,\sigma} \\&+ \Delta \sum\limits_{\langle i,j\rangle} ( {f_2}^\dagger_{i,\uparrow}{f_2}^\dagger_{j,\downarrow}-{f_2}^\dagger_{i,\downarrow}{f_2}^\dagger_{j,\uparrow}+{\rm h.c.}) \label{eq:mfhamf2}
     \end{split}
    \end{equation}

As the energetics of this layer are completely decoupled from the other two, we can choose $t_{f_2}=1$ as the reference. This Hamiltonian allows us to smoothly interpolate between a zero- and $\pi$-flux spin liquid on the second ancilla layer by tuning the pairing $\Delta$ from $0$ to $t_{f_2}$~\cite{LeeRMP}. While we always allow for $\Delta$ to act as a variational parameter, we will also consider these two special cases separately, as it can be quite hard to hit exact values during optimization.

The variational {\it Ansatz} for the wave-function is now given by~\cite{zhang2020}
\begin{equation}
    \ket{\Psi} = \mathcal{P}_S \mathcal{P}_{G,1}\mathcal{P}_{G,2} \ket{c,f_1}\ket{f_2}, \label{eq:groundstate}
\end{equation}
where $\ket{c,f_1}$ and $\ket{f_2}$ are the Slater determinant ground states of \eqref{eq:mfham} and \eqref{eq:mfhamf2}, respectively. $\mathcal{P}_{G,1/2}$ are Gutzwiller projectors, enforcing the half-filling per site constraint \eqref{eq:spinconst} rendering the spin-to-fermion mapping exact. $\mathcal{P}_{S}$ subsequently projects the ancilla qubit layers onto local rung-singlets, reminiscent of an infinitely strong local antiferromagnetic coupling between the layers. This effectively screens out the magnetic background and, at the same time, transfers properties of the spin liquid on the second layer to the itinerant fermions. We emphasize that the state $\ket{\Psi}$ is a function only of the positions and spins of the electrons in the physical Hubbard layer.

In this work, we relax $\mathcal{P}_S$: Instead of projecting onto the singlet with $({\mathbf S}_1+{\mathbf S}_2)^2 = 0$, we focus on the subspace of states fulfilling $S^{z}_{1} + S^{z}_{2} = 0$, i.e., in addition to the singlets we also include the $S^{z}=0$ component of the triplet states. This replaces the fully entangled pure state \eqref{eq:groundstate} by a mixed state consisting of all possible configurations fulfilling $S^{z} = 0$ per site \cite{shcakeltonprivate}. We, however, argue that this mostly affects physical expectation values that are off-diagonal in the itinerant configuration, like kinetic terms in the Hamiltonian. In a mixed state, for a given itinerant configuration, the weight of singlet and triplet components in the $S^{z}_{1} + S^{z}_{2} = 0$ subspace is always equal, as $\ket{\uparrow\downarrow}$ consists equally of these contributions. Therefore, optimizing the overlap of the variational state to be maximal with the true ground state, the relative weights of the products of itinerant and $S^{z}=0$ states will be chosen such that the singlet content matches best the true state, while the triplet part, being orthogonal, will not give any contributions to this overlap. Due to normalization, operators diagonal in the itinerant configuration will only connect the same states, collecting contributions according to the singlet weights. Off-diagonal operators, which change the itinerant electronic configuration, on the other hand, will connect differently weighted combinations of $S^{z}=0$ states, therefore being possibly prone to large corrections from the relaxed projection. Since, we are mostly interested on local (hole-)spin-spin correlators, which leave local occupations in the itinerant layer untouched, we expect the relaxed constraint {\it Ansatz} to perform quite well.

\begin{figure*}
    \includegraphics[width=\textwidth]{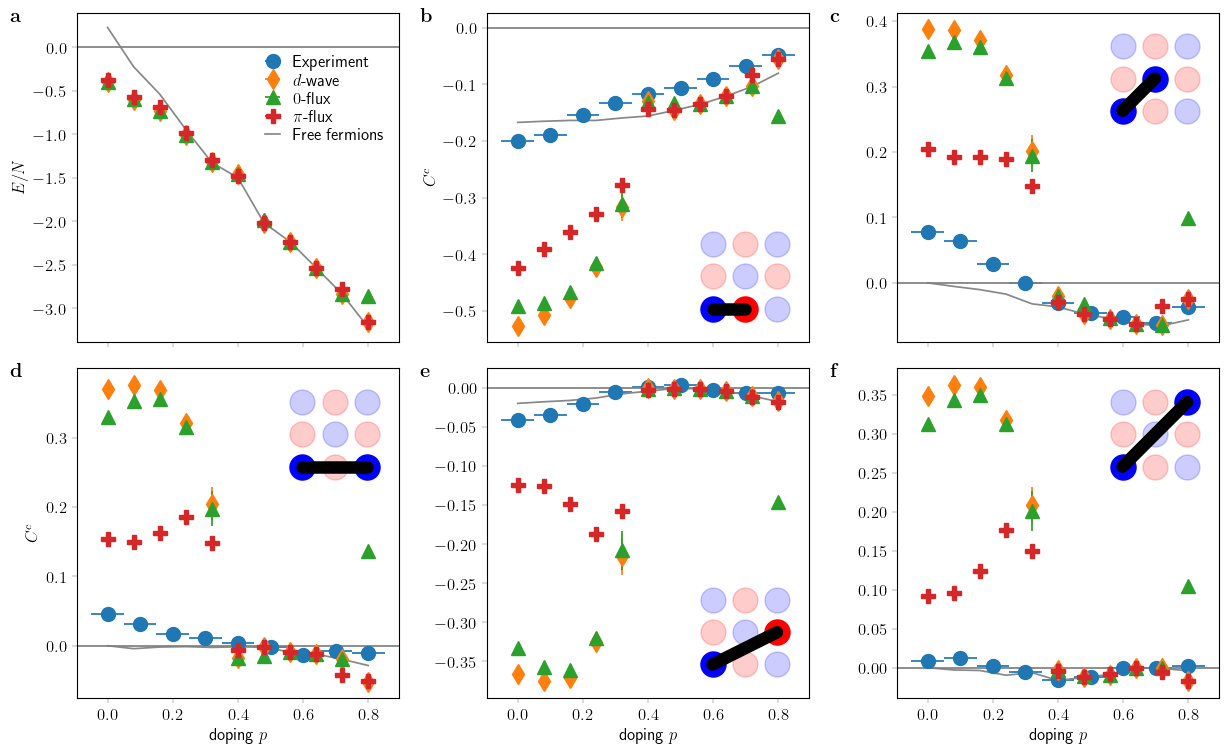}
    \caption{\textbf{a} Doping dependence of the variational energies of the general $d$-wave, $0$-flux and $\pi$-flux {\it Ansätze} in comparison to pure free fermions. For low doping, the ancilla wave-functions clearly outperform free-fermion. \textbf{b}-\textbf{f} Connected spin-spin correlators $C^c$ for various distances as shown in the insets, where blue (red) points indicate (anti-)ferromagnetic correlations at half-filling. We compare our ground-state data for $U=7.4$ to both free fermions and experimental correlators~\cite{koepsell2021}, which have been recorded for $U=8.9$ and $T=0.43$ and find qualitative agreement with the latter. \label{fig:corell}}
\end{figure*}

\begin{figure*}
    \includegraphics[width=\textwidth]{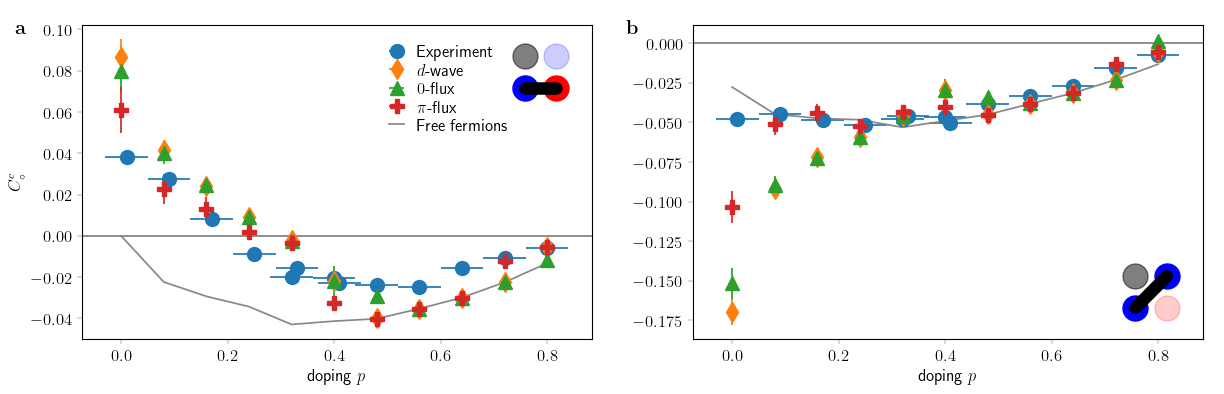}
    \caption{Doping dependence of connected spin-spin correlations in the presence of one hole $C^c_{\circ}$ for \textbf{a} nearest-neighbor and \textbf{b} next-nearest neighbor sites. For the former, we find quantitative agreement with experimental data~\cite{koepsell2021} for moderately low doping, with the general trend of ferromagnetic values being reproduced by all candidate wave-functions, in contrast to free-fermions. On the next-nearest-neighbor level, only the $\pi$-flux {\it Ansatz} quantitatively captures experimental trends away from half-filling.\label{fig:inducedcorell}}
\end{figure*}

{\it Numerical implementation.---} We use  a standard formulation of variational Monte Carlo~\cite{becca2017}, tailored to the trilayer system under investigation. For the Monte Carlo sampling, we implement the Gutzwiller projectors by visiting only configurations which are half-filled per spin. Additionally, the $S^{z}=0$ projection is likewise implemented by only considering micro-states, where spins in the two ancilla layers are oriented anti-parallel per site. At fixed filling, this allows only for a restricted set of updates: in the itinerant layer, we use hopping of electrons to the neighboring sites, while for the ancilla layers, we implement a combined spin swap between two points in both layers simultaneously to keep them anti-parallel. Additionally, we find that concurrent swapping of the configuration between three points of all three layers leads to an increase of the acceptance rate. As the itinerant and first ancilla layer are connected by the hybridization $\Phi$, this is not totally unexpected.

To avoid large degeneracy in the ground state of the mean-field Hamiltonians, we choose specific twisted boundary conditions, adding a phase $ e^{\dot\i \phi}$ when crossing the boundary. In the $x$-direction, we choose $\phi = \pi/2$, while we use anti-periodic boundary conditions ($\phi = \pi$) in the y-direction. This effectively shifts the reciprocal lattice, such that all energies are only four-fold degenerate due to lattice rotational symmetry and spin degeneracy. This allows us to numerically access fillings in steps of eight holes.

{\it Results.---} We perform optimizations of the variational wave-functions for the $0$ and $\pi$-flux cases, as well as the less constrained full $d$-wave {\it Ansatz} at hole-dopings $p=0$ (half-filling) up to $p=0.8$. In the following comparisons, we show for each {\it Ansatz} the expectations values obtained from the lowest energy wave-function as measured by the physical Hamiltonian \eqref{eq:hubbardham}. For reference, we always show data for a free-fermion wave-function, obtained by setting $\Phi=0$. We will benchmark the performance of our wave-functions against cold-atom quantum simulator data from Ref.~\cite{koepsell2021}. Here, we do not expect full numerical agreement as the experiments are carried out at finite temperature of about $T=0.5t$. Although small, this will, in general, reduce the magnitude of measured correlators. Close to half-filling, however, there is another effect, when it comes to measurements involving holes: their intrinsic number here is very small, such that thermal activation can trigger larger values, especially for correlations involving multiple holes.

It is known, that, in contrast to the FL* phase for small dopings, the system transitions into a FL for sufficiently large hole-doping, when the electrons are dilute enough, that Hubbard interactions hardly play a role. In this regime, we expect our wave-functions to perform extremely well, as this edge case is exactly captured by our construction. This is already visible in the energy-dependence of the optimized wave-functions in Fig.~\ref{fig:corell}(a). For small dopings, all {\it Ansätze} outperform the pure free-fermion case, while starting from $p\geq0.4$ the variational energies start to coincide with the free-fermion ones. At all $p$, the $0$-flux and $d$-wave states slightly outperform the $\pi$-flux energetically, but as explicated above, the energy is the measurement we expect to be modified most drastically by the relaxation of the singlet constraint.

To get more reliable results, we turn to connected spin-spin correlators, given by
\begin{equation}
    C^c(\mathbf{r}_1, \mathbf{r}_2) = \eta\left(\ev{S^z_{\mathbf{r}_1}S^z_{\mathbf{r}_1}}-\ev{S^z_{\mathbf{r}_1}}\ev{S^z_{\mathbf{r}_2}}\right),
\end{equation}
where where the normalization $\eta = 1/(\sigma(S^z_{\mathbf{r}_2})\sigma(S^z_{\mathbf{r}_2}))$ is given by the standard deviations of the spin operators. 

Our numerical results shown in Figs. \ref{fig:corell}(b)-(f) demonstrate good qualitative agreement with the experimental data. For $p<0.4$, we find, that the optimized wave-functions all significantly overestimate the spin correlations at all distances. It is worth noting however that the experiment was conducted for $U=8.9$, differing from the value of $U=7.4$ we employ for variational optimization, to be in agreement with the other experimental data. This increase in interaction will, however, weaken the correlations in addition to the finite temperature in the experiment. This can be rationalized by considering a half-filled Hubbard model in the strong coupling limit. Taking into account the hopping $t$ perturbatively, leads to an antiferromagnetic Heisenberg interaction of strength $J=-t^2/U$. As the correlation strength will be proportional to $|J|$, an increase in $U$ leads to weaker correlations. This argument is also expected to qualitatively hold for small dopings away from $p=0$.

At half-filling, all variational states show antiferromagnetic/ferromagnetic distance dependent correlations, in accordance with experimental results, therefore vastly improving over a free-fermion wave-function, which would predict vanishing magnetic correlations for the ferromagnetic second, third and fifth nearest neighbors. With increasing hole-doping, $C^c$ at these distances shows a sign change around $p=0.4$, which is captured by the variational states, although the transition is more pronounced there. For $p>0.4$, both experimental data and variational states start to agree with a free-fermion {\it Ansatz}, therefore we find excellent quantitative agreement for these dopings. Comparing the three cases of variational states we use, we find, that the $\pi$-flux performs best with about $<50\%$ deviation from the experimental values at small dopings. Although energetically almost on par, the $0$-flux {\it Ansatz} slightly outperforms the general $d$-wave state. 

To shed more light on the physical performance of the variational states, we now turn to polaronic correlations. In the weakly doped regime, holes are expected to be surrounded by a magnetic perturbation cloud, changing the spin-spin correlators around them. To capture this effect, we measure the connected hole-spin-spin correlator
\begin{equation}
    \begin{split}
    C^c_{\circ}(\mathbf{r}_1,\mathbf{r}_2,\mathbf{r}_3) = \eta&(\ev{h_{\mathbf{r}_3}S^z_{\mathbf{r}_1}S^z_{\mathbf{r}_2}} - \ev{h_{\mathbf{r}_3}}\ev{S^z_{\mathbf{r}_1}S^z_{\mathbf{r}_2}} \\&- \ev{S^z_{\mathbf{r}_1}}\ev{h_{\mathbf{r}_3}S^z_{\mathbf{r}_2}} - \ev{S^z_{\mathbf{r}_2}}\ev{h_{\mathbf{r}_3}S^z_{\mathbf{r}_1}} \\&+2 \ev{h_{\mathbf{r}_3}}\ev{S^z_{\mathbf{r}_1}}\ev{S^z_{\mathbf{r}_2}}).
    \end{split}
\end{equation}
The numerical results in Fig.~\ref{fig:inducedcorell} show excellent quantitative agreement with the experimental cold-atom data. Most importantly, for small hole-dopings the polaronic dressing cloud around the hole changes the sign of nearest-neighbor spin correlations to ferromagnetic, compared to the bare antiferromagnetic correlations discussed above. With increasing doping the system evolves towards the non-interacting FL limit, leading to restoration of AF correlations at around $p\sim 0.3$, in accordance with experimental observations. This behavior is in contrast to pure free-fermion FL behavior, which would show AF $C^c_{\circ}$ for all dopings. 

\begin{figure*}
    \includegraphics[width=\textwidth]{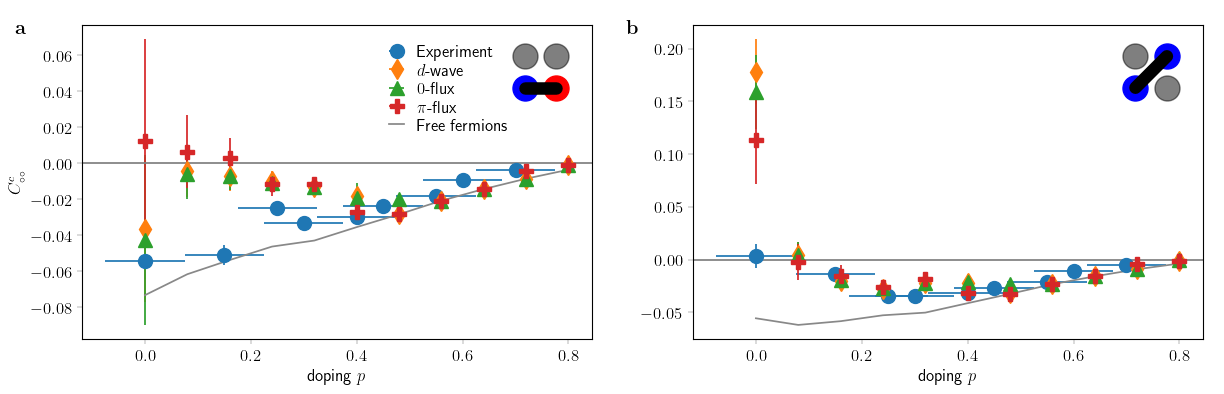}
    \caption{Evolution of spin-spin correlators in the vicinity of two holes $C^c_{\circ\circ}$ with hole-doping. For the nearest-neighbor correlator \textbf{a}, the $\pi$-flux {\it Ansatz} fails to reproduce the antiferromagnetic tendency close to half-filling of the cold-atom experiment~\cite{koepsell2021}, while the other {\it Ansätze} qualitatively capture this feature, which is best described by pure free-fermion behavior. For the next-nearest neighbor correlations \textbf{b}, in contrast, the variational wave-function achieves quantitative agreement away from half-filling, which pure free-fermions fail to do.\label{fig:doubleinducedcorell}}
\end{figure*}

Similarly, second nearest-neighbor polaronic correlations being antiferromagnetic exhibit a sign flipped behavior with respect to their unperturbed counterparts. This, however, is already captured by a pure FL, with slight enhanced strength at low dopings. For half-filling, all trial wave-functions overestimate the experimental values of $C^c_{\circ}$, while for $p=0.8$ the $\pi$-flux state already quantitatively captures both nearest- and second nearest-neighbor polaronic correlations. Both $0$-flux and $d$-wave {\it Ansatz} show stronger correlations compared to the cold-atom setup, however, this is to be expected due to finite temperature in experiments. Please note, that even in the FL regime ($p>0.4$), experimental correlations are reduced compared to free-fermion ones, what we again attribute to finite temperature effects.

As a third signature of FL* physics, we present spin-spin correlations in the vicinity of two neighboring holes, given by
\begin{equation}
    \begin{split}
        C^c_{\circ\circ}(\mathbf{r}_1,\mathbf{r}_2&,\mathbf{r}_3,\mathbf{r}_3) = \eta\big(\ev{h_{\mathbf{r}_4}h_{\mathbf{r}_3}S^z_{\mathbf{r}_1}S^z_{\mathbf{r}_2}} 
        - \ev{h_{\mathbf{r}_4}} \ev{h_{\mathbf{r}_3}S^z_{\mathbf{r}_2}S^z_{\mathbf{r}_1}}\\
        &- \ev{h_{\mathbf{r}_3}} \ev{h_{\mathbf{r}_4}S^z_{\mathbf{r}_2}S^z_{\mathbf{r}_1}} 
        - \ev{h_{\mathbf{r}_4}h_{\mathbf{r}_3}} \ev{S^z_{\mathbf{r}_2}S^z_{\mathbf{r}_1}}\\
        &- \ev{S^z_{\mathbf{r}_2}} \ev{h_{\mathbf{r}_4} h_{\mathbf{r}_3}S^z_{\mathbf{r}_1}}
        - \ev{S^z_{\mathbf{r}_1}} \ev{h_{\mathbf{r}_4} h_{\mathbf{r}_3}S^z_{\mathbf{r}_2}}\\
        &- \ev{h_{\mathbf{r}_4} S^z_{\mathbf{r}_2}} \ev{ h_{\mathbf{r}_3}S^z_{\mathbf{r}_1}}
        - \ev{h_{\mathbf{r}_4} S^z_{\mathbf{r}_1}} \ev{ h_{\mathbf{r}_3}S^z_{\mathbf{r}_2}}\\
        &+ 2\ev{h_{\mathbf{r}_4}} \ev{h_{\mathbf{r}_3}} \ev{S^z_{\mathbf{r}_2}S^z_{\mathbf{r}_1}}
        +2 \ev{h_{\mathbf{r}_4} h_{\mathbf{r}_3}} \ev{S^z_{\mathbf{r}_2}}\ev{S^z_{\mathbf{r}_1}}\\
        &+2 \ev{h_{\mathbf{r}_4} S^z_{\mathbf{r}_2}} \ev{ h_{\mathbf{r}_3}}\ev{S^z_{\mathbf{r}_1}}
        +2 \ev{h_{\mathbf{r}_4} S^z_{\mathbf{r}_1}} \ev{ h_{\mathbf{r}_3}}\ev{S^z_{\mathbf{r}_2}}\\
        &+2 \ev{h_{\mathbf{r}_3} S^z_{\mathbf{r}_2}} \ev{ h_{\mathbf{r}_4}}\ev{S^z_{\mathbf{r}_1}}
        +2 \ev{h_{\mathbf{r}_3} S^z_{\mathbf{r}_1}} \ev{ h_{\mathbf{r}_4}}\ev{S^z_{\mathbf{r}_2}}\\
        &-6 \ev{h_{\mathbf{r}_4}} \ev{h_{\mathbf{r}_3}} \ev{S^z_{\mathbf{r}_2}} \ev{S^z_{\mathbf{r}_1}}\big)\, .
    \end{split}
\end{equation}
In this setup, the polaronic dressing clouds overlap and cancel their respective effects. In the case of nearest-neighbor holes, Fig.~\ref{fig:doubleinducedcorell}(a), the nearest-neighbor spin correlators in experiment show an antiferromagnetic behavior, as seen for the bare ones, and mimic closely the values obtained from free fermions. Our variational wave-functions, however, although showing the right sign of correlations, underestimate this correlator in the low doping regime up to the crossover to the FL. Here, especially the $\pi$-flux state fails to reproduce the right correlations, even yielding FM expectation values, however, the error-bars are quite large, owing to the low number of fluctuation induced hole pairs in this case.

For next-nearest neighbor holes neighboring a next nearest-neighbor spin correlator, the picture is completely opposite. Although ferromagnetic in the unperturbed case, the two neighboring polaronic perturbations close to half-filling drive the experimental values to be uncorrelated within error-bars, with a slight tendency to ferromagnetism. At half-filling, the variational wave-functions severely overestimate this tendency, but already at $p=0.08$, we also find vanishing values of $C^c_{\circ\circ}$ within statistical errors. For higher dopings, the variational data almost perfectly reproduce the experiment.

We explain these observations as follows: fluctuation induced hole pairs, which are always accompanied by a doublon pair behave in a qualitatively different way from doping induced hole pairs, which do not necessitate doubly filled sites in their vicinity. In the nearest-neighbor case, the fluctuation induced hole pairs lead to antiferromagnetic correlations, as can be seen from the free-fermion data at half-filling. Apparently, the doping induced hole-pairs, which exist away form $p=0$ contribute a ferromagnetic component to the nearest-neighbor $C^c_{\circ\circ}$. As, in contrast to the experiment, the variational calculations measure strictly at $T=0$, mainly holes originating from doping will contribute to these expectation values, in contrast to the thermally activated ones in cold-atom data~\cite{koepsell2021}. 

For the next-nearest-neighbor $C^c_{\circ\circ}$, it is known, that doublon-hole fluctuations lead to a ferromagnetic contribution~\cite{koepsell2021}, which at half-filling seems to be overestimated by the variational {\it Ansätze}, while this levels out once doping induced hole take over.

{\it Conclusion.---} We have demonstrated the ability and scope of ancilla qubit wave-functions to accurately describe the magnetic polaron correlations in the fractionalized Fermi liquid suggested to account for the pseudo-gap metal phase at low dopings. Employing three different spin liquid {\it Ansätze} ($0$-flux, $\pi$-flux, and $d$-wave) as a hidden layer in the ancilla construction provides us with the needed versatility of reproducing the cold-atom simulator data. In general, all three wave-functions are capable of capturing the qualitative behavior of bare spin-spin correlations, with quantitative differences readily explained by experimental temperature and interaction parameters which are not accounted for in the variational ground state calculations. Polaronic correlators show quantitative agreement away from half-filling for all {\it Ansätze}. The $\pi$-flux state, however, fails to reproduce the sign of nearest-neighbor correlators under the influence of two holes, suggesting it to be less competitive compared to the other two. The two-hole induced second-nearest neighbor correlations are captured well by all {\it Ansätze}.

In summary, the ancilla qubit wave-functions enable the description of the highly correlated FL* phase with a low-parametric trial state, i.e., in our case, only one and two parameters for $0/\pi$-flux and $d$-wave, respectively. Studying more realistic models of, e.g., cuprates to test the scope of this family of ground-state {\it Ansätze} suggests itself for future endeavors. From a methodological viewpoint, the flexibility in the spin liquid {\it Ansatz} implies that the description in terms of ancilla qubit wave functions is possibly not just restricted to the FL* phase, but also that additional phenomenological constraints might be desirable to uniquely pin down a single most appropriate {\it Ansatz}.

{\it Acknowledgments.---} We thank Immanuel Bloch, Eugene Demler, Fabian Grusdt, Henry Shackleton, Shiwei Zhang, and Ya-Hui Zhang for inspiring discussions. The work of R.T. and T.M. is funded by the Deutsche Forschungsgemeinschaft (DFG, German Research Foundation) through Project-ID 258499086 - SFB 1170, the W\"urzburg-Dresden Cluster of Excellence on Complexity and Topology in Quantum Matter – ct.qmat Project-ID 390858490 - EXC 2147. S.S. is supported by the U.S. National Science Foundation grant No. DMR-2245246 and by the Simons Collaboration on Ultra-Quantum Matter which is a grant from the Simons Foundation (651440, S.S.). The work of Y.~I.\ was performed, in part, at the Aspen Center for Physics, which is supported by National Science Foundation Grant No.~PHY-2210452 and a grant from the Simons Foundation (1161654, Troyer). This research was supported in part by grant NSF PHY-2309135 to the Kavli Institute for Theoretical Physics (KITP). Y.~I.\ acknowledges support from the ICTP through the Associates Programme, from the Simons Foundation through Grant No.~284558FY19, and IIT Madras through the Institute of Eminence (IoE) program for establishing QuCenDiEM (Project No.~SP22231244CPETWOQCDHOC). Y.~I.\ also acknowledges the use of the computing resources at HPCE, IIT Madras.

{\it Note added.---} We mention an independent simultaneous work~\cite{Henry-2024} that jointly appears on the arXiv.

\end{document}